\PassOptionsToPackage{square, comma, sort&compress}{natbib}
\documentclass[preprint,12pt]{elsarticle}

\usepackage{amsmath}
\usepackage[colorlinks=true,linkcolor=blue,urlcolor=blue,citecolor=blue]{hyperref}
\usepackage{graphicx}
\usepackage[squaren, Gray, cdot]{SIunits}
\usepackage{amssymb}

\usepackage{tikz}

\begin{document}

\begin{frontmatter}

\title{Coherent Optical Processes on Cs D$_2$ line Magnetically Induced Transitions}

\author[IPR]{Armen Sargsyan}
\author[IPR]{Arevik Amiryan}
\author[IPR]{Ara Tonoyan}
\author[JGU]{Emmanuel Klinger\corref{mycorrespondingauthor}}
\cortext[mycorrespondingauthor]{Corresponding author}
\ead{eklinger@uni-mainz.de}
\author[IPR]{David Sarkisyan}

\address[IPR]{Institute for Physical Research – National Academy of Sciences of Armenia, 0203 Ashtarak-2, Armenia}
\address[JGU]{Helmholtz-Institut Mainz -- GSI Helmholtzzentrum f{\"u}r Schwerionenforschung, Johannes Gutenberg-Universit{\"a}t, D-55128 Mainz, Germany}

\begin{abstract}

The increased spectral resolution allowed by the use of extremely thin vapor cells has led to the observation of interesting behaviour of alkali transitions when placed in a magnetic field. Particularly, transitions obeying an apparent $F_e-F_g\equiv\Delta F =\pm2$ selection rule, referred to as magnetically-induced (MI) transitions, have their probabilities largely increase in the intermediate interaction regime while being null at zero and higher magnetic fields. With an 800\,nm-thick Cs vapor cell placed in a field up to 1.5\,kG, we show here that the generation of electromagnetically induced transparency (EIT), realized in $\Lambda$-systems involving $\Delta F =- 2$ MI transitions, is only possible when both the coupling and probe beams are $\sigma^-$-circular polarized, demonstrating that EIT is affected by magnetic circular dichroism. A similar rule of thumb can be extrapolated for $\Delta F =+2$ MI transitions and $\sigma^+$ polarization.  Because of the high frequency shift slope (typ. 4\,MHz/G), the generation of EIT resonances involving MI transitions is interesting, especially in the context of growing attention towards micro-machined alkali vapor cell sensors.
\end{abstract}


\end{frontmatter}


\section{Introduction}
With the advent of extremely-thin-cells, so called nanocells (NC), allowing one to record sub-Doppler resonances while staying in the weak-probe regime of interaction, the interest in studying the evolution of alkali transitions in a magnetic field has been regained. Thanks to the increased frequency resolution, one is able to follow the frequency shift and amplitude evolution of individual transitions as a function of the magnetic field. Such studies were previously extremely complicated to perform with techniques such as saturated absorption spectroscopy which create additional (crossover) resonances increasing the difficulty in interpreting observed spectra \cite{smith2004role,scottoPRA2015}. 

Magnetically-induced (MI) transitions, obeying the apparent selection rule $F_e-F_g\equiv\Delta F=\pm2$, are forbidden at zero magnetic field. For alkali atoms (Cs, Rb, K, Na), MI transitions represent a large class consisting of about 100 atomic transitions exhibiting interesting and important specific features \cite{tremblayPRA1990, sargsyanApplPhysLet2008, sargsyanLPL2014, scottoThesis2016, tonoyanEPL2018, sargsyanJetpLett2021, sargsyanPhysLetA2021}. Interest in MI transitions is mainly due to the fact that their probabilities in a certain range of magnetic field can considerably exceed probabilities of conventional atomic transitions allowed in the absence of magnetic field. The significant change in transition probabilities, in particular the giant increase in the probabilities of MI transitions, is caused by the mixing of magnetic sub-levels because of an external magnetic field \cite{auzinshBook2010}. In addition, thanks to their large frequency shift slope (typically 4\,MHz/G), nearly 4 times larger than conventional transitions \cite{sarkisyanJETP2020}, MI transitions are interesting for field using the wide tunability of alkali transitions in a magnetic field, notable applications include atomic Faraday filters \cite{zielinska2012ultranarrow,keaveney2018optimized}.

Electromagnetically induced transparency (EIT) is a nonlinear phenome-
non where coherence between atomic states is created by a strong laser field (referred to as coupling in the following), creating a transparency window in the absorption spectrum of a weaker laser field (referred to as probe). In conjunction with transparency, a rapid change of refractive index is also produced, which is the core of slow and stored light experiments \cite{fleischhauerRevModPhys77,novikova2012electromagnetically} aiming at quantum memory. Other applications of EIT include electromagnetic field sensing \cite{yudin2010vector,belfi2007cesium,PhysRevA.100.063427,holloway2021electromagnetically} and atomic clocks \cite{vanier2005atomic,hafiz2020protocol}. 

From theoretical calculations, the following rule of thumb was inferred and verified experimentally for MI transitions: transitions following the selection rule $\Delta F =\pm2$ have much larger amplitudes for $\sigma^\pm$ polarized probe light, respectively. The difference of intensities for $\sigma^+$ and $\sigma^-$ radiations can reach several orders of magnitude, thus leading to anomalous circular dichroism \cite{tonoyanEPL2018}. Recently, EIT resonances where formed using $^{87}$Rb \cite{sargsyanOptlett2019} and $^{85}$Rb \cite{sargsyanJETP2021} $\Delta F = +2$ MI transitions along with conventional transitions, when the probe laser is scanned in the vicinity of MI transitions while the couple beam is tuned on conventional transitions. It was shown that EIT resonances are absent when the probe is $\sigma^-$ circular polarized.

Here, we study the formation of EIT resonances when the couple beam is tuned on $\Delta F = -2$ transitions of the Cs D$_2$ line while the probe beam is tuned on conventional transitions, several GHz away. The aim of the present experiment is to reveal which circular polarization combination of the coupling and probe radiation beams ensures the formation of the EIT resonance  with the largest contrast. In Sec.\,\ref{sec:theory}, we recall the calculated probabilities and shifts of transitions in the vicinity of the $F_g = 3 \rightarrow F_e = 2',3',4'$ manifold in the presence of a magnetic field. In Sec.\,\ref{sec:experiment}, we describe the experimental apparatus and study the formation of EIT resonance on MI transitions.

\begin{figure}[h!]
    \centering
   \includegraphics[scale=1]{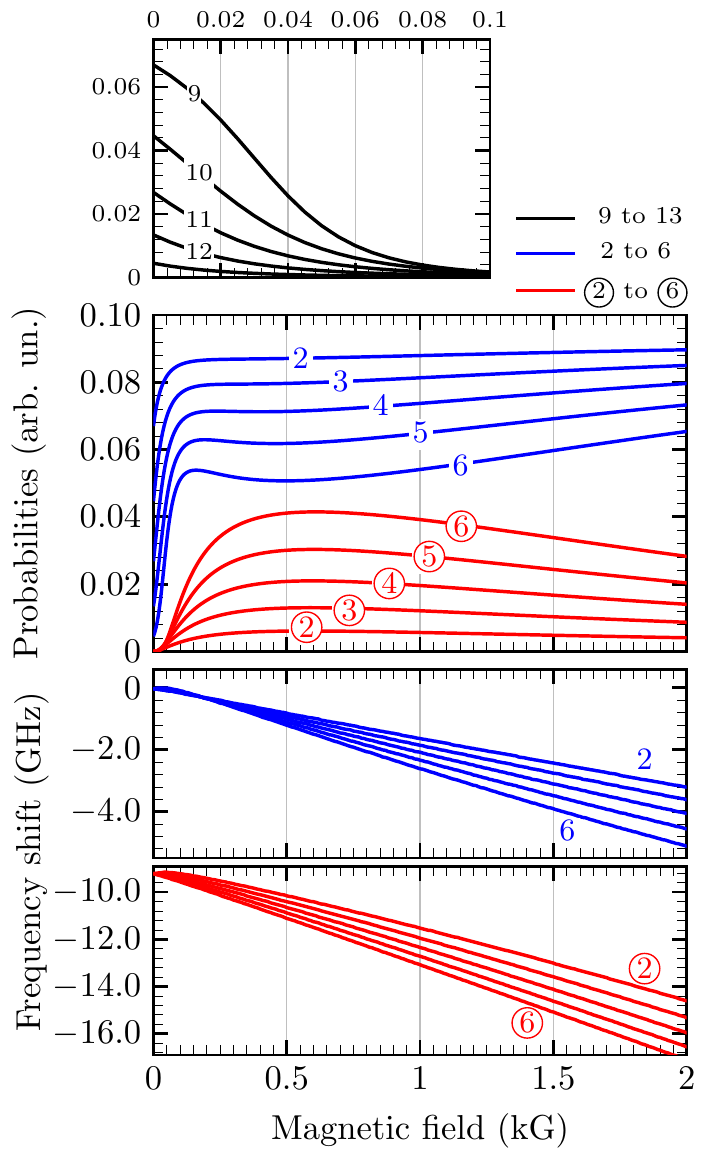}
    \caption{Cesium D$_2$ line transitions in a magnetic field.  The top panel shows the probabilities of $\sigma^-$ transitions $|3,m_F\rangle \rightarrow |2',m_F-1\rangle$, where $m_F=3,2,...,-1$ (labelled 2 to 6, respectively) and $|4,m_F\rangle \rightarrow |2',m_F-1\rangle$, where $m_F=3,2,...,-1$ (labelled respectively \textcircled{2} to \textcircled{6}) as a function of the magnetic field. The middle and bottom panels show the frequency shifts of transitions 2 to 6 (blue) and \textcircled{2} to \textcircled{6} (red) as a function of magnetic field, where the zero frequency is set to that of the unperturbed $3 \rightarrow  2'$ transition. On these two panels, only extreme transitions have been marked to preserve readability.  The top inset shows the probabilities of $\sigma^+$ transitions 9 to 13, occurring between the sublevels $|3,m_F\rangle \rightarrow |2',m_F+1\rangle$ where $m_F=-3,-2,...,1$ respectively; they rapidly tend to zero.}
    \label{fig:shift-proba}
\end{figure}

\section{Cs D$_2$ line in a magnetic field}
\label{sec:theory}

The probabilities and frequency shifts of Cs D$_2$ line ($\sigma^-$) transitions $F_g = 3 \rightarrow F_e = 2'$ (labelled 2 to 6, in blue) and $F_g = 4 \rightarrow F_e = 2'$ (labelled \textcircled{2} to \textcircled{6}, in red) as a function of the magnetic field are depicted in Fig.\,\ref{fig:shift-proba}. We chose to label transitions according to their initial $|F, m_F\rangle$ (ground states) and final $|F', m'_F\rangle$ (excited states) quantum numbers, i.e. in the coupled basis of states. As seen, the transition probabilities grow with increasing field while experiencing a red shift. Because the probabilities of MI transitions \textcircled{2} to \textcircled{6} are high enough in the range of 1 to 2\,kG, one should be able to use them to form EIT resonances. A diagram of transitions for $\sigma^-$ polarized light can be seen in Fig.\,\ref{fig:setup}. The probabilities of transitions 9 to 13, equivalent of 2 to 6 but with $\sigma^+$ polarization, are shown in the top inset. As can be seen, for $B> 100$\,G, their probabilities tend to zero; therefore, these transitions cannot be used in large magnetic fields. This fact is very important for understanding why EIT resonances cannot be formed with other polarization than $\sigma^-$.
As demonstrated in \cite{tonoyanEPL2018}, the probabilities of transitions $4 \rightarrow 2'$ with $\sigma^+$ polarisation are lower than that for $\sigma^-$ polarisation by several orders of magnitude. 
 
 These calculations have been performed using the theoretical model presented in \cite{tremblayPRA1990,tonoyanEPL2018,sargsyanPhysLetA2021}, which describes the change in the probabilities and frequencies of atomic transitions in the magnetic field using the Hamiltonian matrix including all transitions of the hyperfine manifold.

\section{Experiment}
\label{sec:experiment}

\subsection{Experimental setup}
\begin{figure}[h!]
    \centering
    \includegraphics[width=0.75\textwidth]{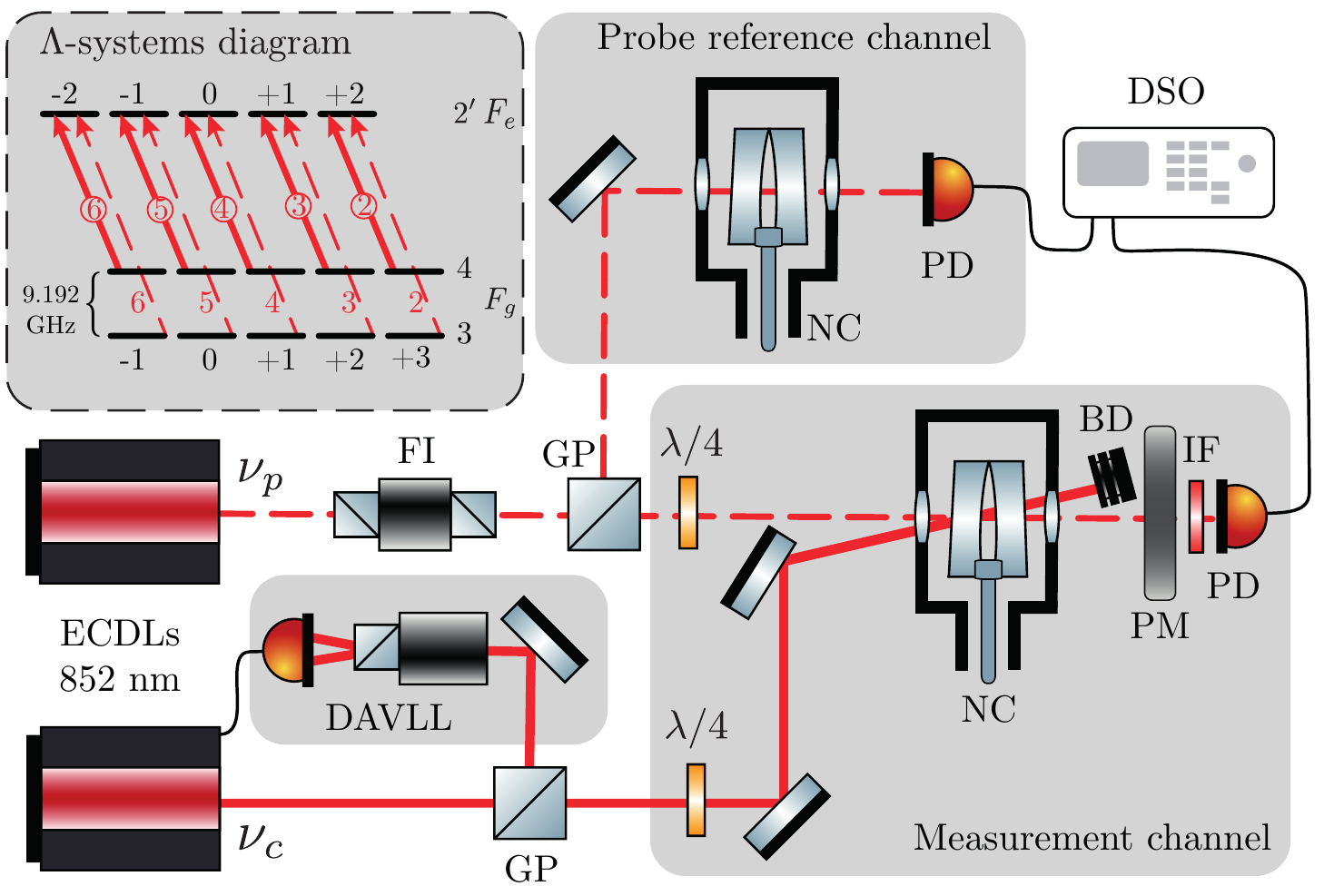}
    \caption{Layout of the experimental setup. ECDLs -- Extended cavity diode lasers, FI -- Faraday insulator, GP -- Glan polarisers, $\lambda/4$ -- quarter-wave plate, NC -- Cs-filled nanocell in a furnace, BD -- beam dump, PM -- permanent magnet, IF -- interference filter, PD -- photodiode, DSO -- digital storage oscilloscope. The angle between the probe ($\nu_p$) and the coupling ($\nu_c$) beams is about 20\,mrad. The inset shows the transitions forming $\Lambda$-systems on the Cs D$_2$ line, with dashed lines for probe beam and solid lines for coupling beam.}
    \label{fig:setup}
\end{figure}

The layout of the experimental setup is shown on Fig.\,\ref{fig:setup}. Two extended cavity diode lasers are tuned in the vicinity of the Cs D$_2$ line, with a wavelength $\lambda\approx 852 \,\rm{nm}$. Precisely, $\Lambda$-systems are formed on this manifold by scanning the frequency $\nu_p$ of a VitaWave laser ($\delta\nu_p \approx 1\, \rm{MHz}$) \cite{vassilievRevSciInstrum2006} in the vicinity of the $F_g=3 \rightarrow F_e =2'$ manifold  while keeping the frequency $\nu_c$  from a MOGLabs “cateye” laser ($\delta \nu_c \approx 100\, \rm{kHz}$) in resonance with one of the $ 4 \rightarrow 2'$ MI transitions, see the inset in Fig.\,\ref{fig:setup}. A fraction about 10\% of the coupling radiation power was sent to a frequency stabilisation unit based on the DAVLL method \cite{sargsyan2009efficient}.

As we aim at revealing which circular polarisation combination of the coupling and probe radiation beams ensures the formation of the EIT resonance  with the largest contrast, it was necessary to vary these beams' polarisation independently. This was ensured with a noncollinear geometry for the coupling and probe radiations as shown in Fig.\,\ref{fig:setup}, where the angle of convergence between the two 1\,mm-diameter beams in the NC is estimated to about 20 mrad. The polarisation of  probe and coupling radiation was managed by two set or Glan polarisers and quarter-wave plates. Both beams were illuminating the NC on the thickness $\ell=\lambda$. The light in the measurement channel and reference channel was detected using FD-24K photodiodes, later amplified and fed to a four-channel digital oscilloscope Tektronix TDS2014B. The contribution of stray light in the measurement channel was removed by placing a 10\,nm-wide pass-band interference filter in front of the photodiode. Additionally, the coupling radiation was suppressed after the NC using a beam dump. The powers $P_c$ and $P_p$ of the coupling and probe radiation beams were varied in the range of 20--40 and 0.1--0.2\,mW, respectively, using neutral density filters (not shown on Fig.\,\ref{fig:setup}). A fraction of the probe radiation power was guided to a frequency reference channel, containing an additional Cs NC illuminated on the thickness  $\ell = \lambda/2 \approx 426\,\rm{nm}$ \cite{sargsyanLPL2014}. The reservoir temperature of both NCs was kept at about $110^\circ$C, corresponding to a vapor number density $N \approx 2\times10^{13}\,\rm{cm}^{-3}$.

As previously shown, the use of NC with thicknesses $\ell=\lambda$ or $2\lambda$, where $\lambda$ is the wavelength of the resonant laser radiation, allows the formation of EIT resonances with higher contrast \cite{sargsyanOptlett2019}. The ``technical contrast” is defined as the ratio of EIT-induced change in the absorption (\textit{i.e.} the amplitude of the EIT-resonance) to the peak absorption of the vapors when the coupling radiation is absent \cite{sargsyanJModOpt2015}. Details on the NC construction and characterization can be found in \cite{sargsyanPhysLetA2021}. Thanks to the small size of the vapor column probed by the laser beam, strong permanent magnets can be used. Indeed, despite the fact that the field gradient can reach $150\,\rm{G} / \rm{mm}$ near the surface of typical permanent magnets, the field varies by less than 0.2\,G along the vapor column. It results a spectral broadening of the transitions by less than 1\,MHz comparable, in the worst-case scenario, to the effect of the laser linewidth. In this experiment, magnetic fields were produced by a neodymium–iron–boron permanent magnet calibrated by a Teslameter HT201 magnetometer; the magnet was placed near the back window of the NC and had a small hole for transmission of the probe beam. The magnetic field was varied by varying the distance from the magnet to the window of the NC.

\begin{figure}[ht]
    \centering
    \includegraphics{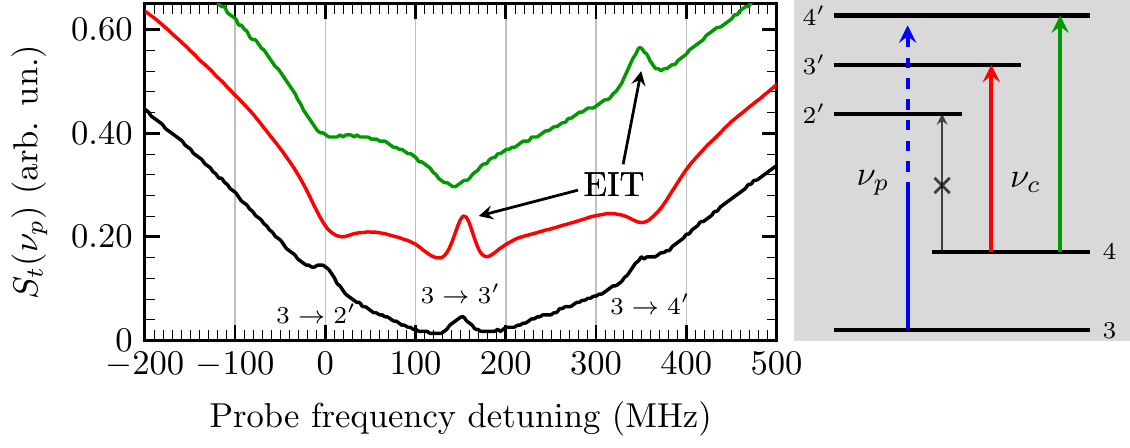}
    \caption{Illustration of the qualitative difference between Velocity Selective Optical Pumping and Electromagnetically Induced Transparency. The curves have been shifted vertically for clarity. The inset shows the $\Lambda$-systems involved in the formation of EIT resonances. As transition $4\rightarrow 2'$ is non-resonant at zero field, it cannot be used to realize EIT and has thus been crossed-out on the diagram. The frequency of transition $3\rightarrow 2'$ was chosen as the zero detuning.}
    \label{fig:VSOP-EIT-Dicke}
\end{figure}
\subsection{Electromagnetically induced transparency on MI transitions}

In \cite{sargsyanApplPhysLet2008}, it was shown that so-called velocity selective optical pumping (VSOP) resonances are formed in the transmission spectrum owing to optical pumping occurring in NCs with $\ell = \lambda$ or $2\lambda$. The spectral width of these resonances are smaller than the Doppler width by a factor of 10 to 20. These resonances demonstrate a decrease in absorption and are located exactly on atomic transition frequencies. Moreover, if the NC is illuminated on the thickness $\ell=\lambda+n\lambda/2$, where $n$ is an integer, then a sub-Doppler resonance is formed, demonstrating increased absorption \cite{andreevapPhysRevA2007}. These two effects should not be mixed with EIT, which demonstrates a decrease in absorption. These phenomena and difference are experimentally illustrated on  Fig.\,\ref{fig:VSOP-EIT-Dicke}. In this figure, we show the transmission spectra $S_t(\nu_p)$ of the probe radiation, in the vicinity of transitions $3\rightarrow 2',3',4'$, recorded at zero field, when the coupling laser frequency is tuned to the transition $4\rightarrow 4'$ (top green line), tuned to $4\rightarrow 3'$ (middle red line).
The bottom black line shows the probe transmission spectrum when the coupling laser is absent; only VSOP resonances can be observed. The corresponding atomic transitions involved in the formation of EIT resonances are shown in the right inset. Note that at zero field, only two EIT resonances can be formed, since transition $4\rightarrow 2'$ is non resonant. However, despite being null at zero magnetic field, probabilities of transitions arising from $4\rightarrow 2'$ are significant for large $B$-field (see Fig.\,\ref{fig:shift-proba}), allowing them to be used to realize EIT. Note that, in Fig.\,\ref{fig:VSOP-EIT-Dicke}, some VSOP resonances of reduced absorption convert to VSOP of increased absorption, see transitions $3\rightarrow2'$ and $3\rightarrow3'$ (green curve) and transitions $3\rightarrow2'$ and $3\rightarrow4'$ (red curve) when EIT resonances are formed on transitions $3\rightarrow 4'$ and $3\rightarrow3'$, respectively. This is a consequence of the coupling laser transferring some population from $F_g=4$ to $F_g=3$ via the excited states.  Since the population in $F_g=3$ increases, the VSOP resonance shows increased absorption, see e.g. \cite{sargsyan2012electromagnetically}.

\begin{figure}[ht]
    \centering
    \includegraphics{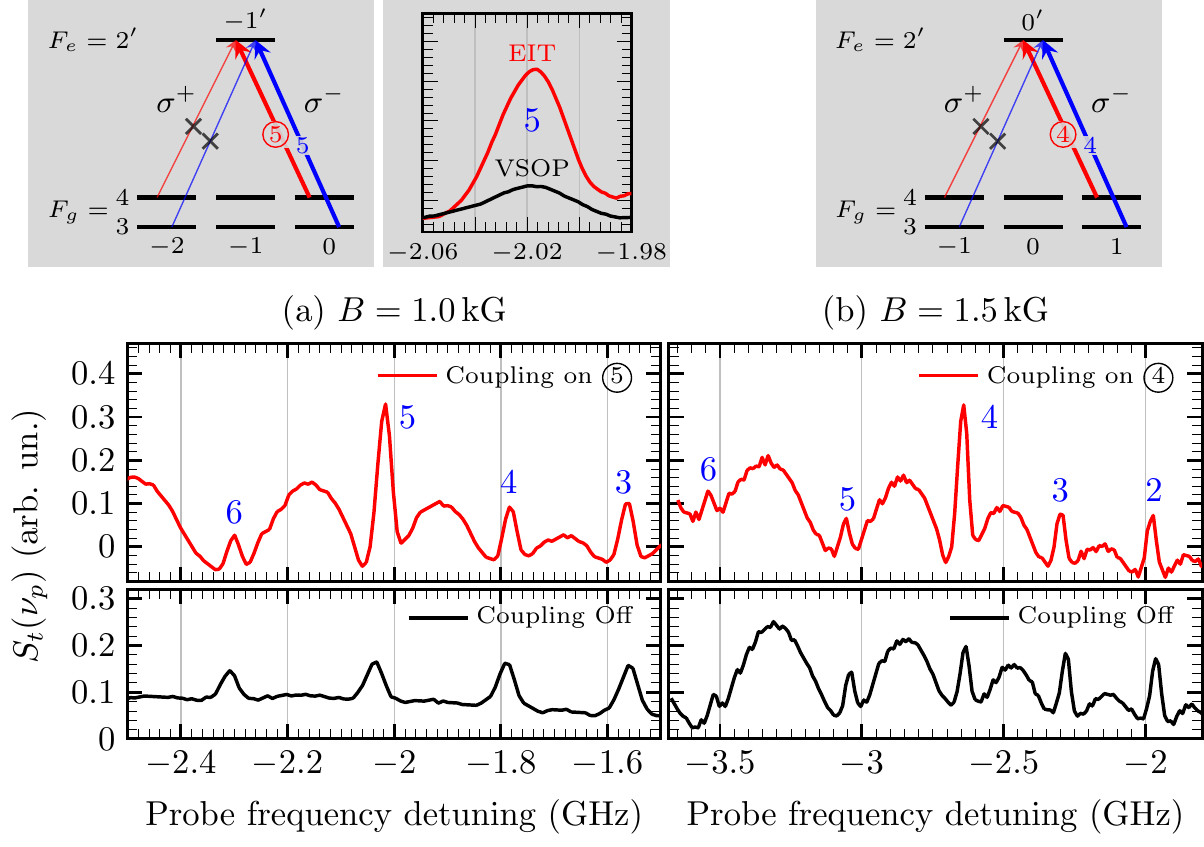}
    \caption{Probe transmission spectra at two intermediate magnetic field values. (a) $B=1.0$\,kG, coupling tuned to \textcircled{5} or switched-off (bottom); (b) $B=1.5$\,kG, coupling tuned to \textcircled{4} (top) or switched-off (bottom). The zero frequency detuning is set to the unperturbed $3 \rightarrow  2'$ transition. The left and right insets show the $\Lambda$-systems diagram, where, for simplicity, only the levels participating in EIT are shown. Probabilities of the transitions $|3, -2\rangle \rightarrow |2', -1'\rangle$ and $|4, -2\rangle \rightarrow |2', -1'\rangle$  ($\sigma^+$ polarization) are close to zero; therefore, they are crossed out. A zoomed area of plot (a), showing the difference between EIT and VSOP resonances, is presented in the middle inset.}
    \label{fig:TR}
\end{figure}

Figure \ref{fig:TR}(a) shows the transmission spectrum of the probe radiation in a longitudinal magnetic field of 1\,kG. In the presence of magnetic field, $\Lambda$-systems involving different $m_F$ magnetic sublevels are formed. Here, the frequency of the probe laser is scanned in the vicinity of the transition $|3, 0\rangle \rightarrow |2', -1'\rangle$, labelled $5$, whereas the frequency of the coupling laser is tuned to the MI transition $|4, 0\rangle \rightarrow |2', -1'\rangle$, labelled \textcircled{5}, see the left inset. The probe spectrum demonstrates higher transmission on transition 5 when the coupling laser is switched-on (top panel), and disappear when switched-off (bottom panel).  No EIT resonances could be formed when any of the probe or coupling radiations was $\sigma^+$-polarized and tuned on $|3, -2\rangle \rightarrow |2', -1'\rangle$ or $|4, -2\rangle \rightarrow |2', -1'\rangle$, respectively. Thus, both the probe and the coupling radiation must have $\sigma^-$ polarization to form an EIT resonance. The right inset in Fig.\,\ref{fig:TR}(a) shows the profile of the EIT and VSOP resonances superimposed on one another. Amplitude of the EIT resonance is seen to be about 4 times larger than that of the VSOP; the spectral width of the EIT resonance (about $30\,\rm{MHz}$) is also 1.3 times smaller which is characteristic of the EIT process observed in NC \cite{sargsyanOL2019, fleischhauerRevModPhys77, kitching2018chip,bhushanPhysRevA2019}.

Figure\,\ref{fig:TR}(b) shows the probe transmission spectra in a longitudinal magnetic field of 1.5\,kG. The probe frequency is scanned in the vicinity of the transition 4 ($|3,1\rangle \rightarrow |2', 0'\rangle$), while the coupling frequency is on resonance with the MI transition $|4,1\rangle \rightarrow |2', 0'\rangle$, labelled \textcircled{4} (top panel) or switched-off (bottom panel). The $\Lambda$-system for the formation of the EIT-resonance using MI transitions \textcircled{4} is shown in the inset, for simplicity only the levels participating in the EIT process are depicted. As the probabilities of transitions $|3, -2\rangle \rightarrow |2', 0'\rangle$ and $|4, -1\rangle \rightarrow |2', 0'\rangle$  ($\sigma^+$ polarization) are close to zero, they have been crossed out on the diagram. The bottom panel shows the transmission spectra in the absence of coupling radiation; only VSOP resonances are observed.

\subsection{Second derivative spectrum}
To further narrow atomic lines, second order derivative (SD) of the transmission spectrum,
was performed, see e.g. \cite{savitzkyAnChem1964, talsky1994, sargsyanOL2019}. This is particularly important for the frequency separation of closely-spaced atomic transitions, especially when a large number of them are present in a small frequency window.  As the double differentiation suppresses any slow variations, the SD spectrum is background free, a feature which is technically convenient. Note that,  because of the nature of the double differentiation, $S''_t$ is multiplied by $-1$ to preserve the same reading as in the transmission spectrum where a higher value means more transmitted (probe) light.

\begin{figure}[h!]
    \centering
    \includegraphics{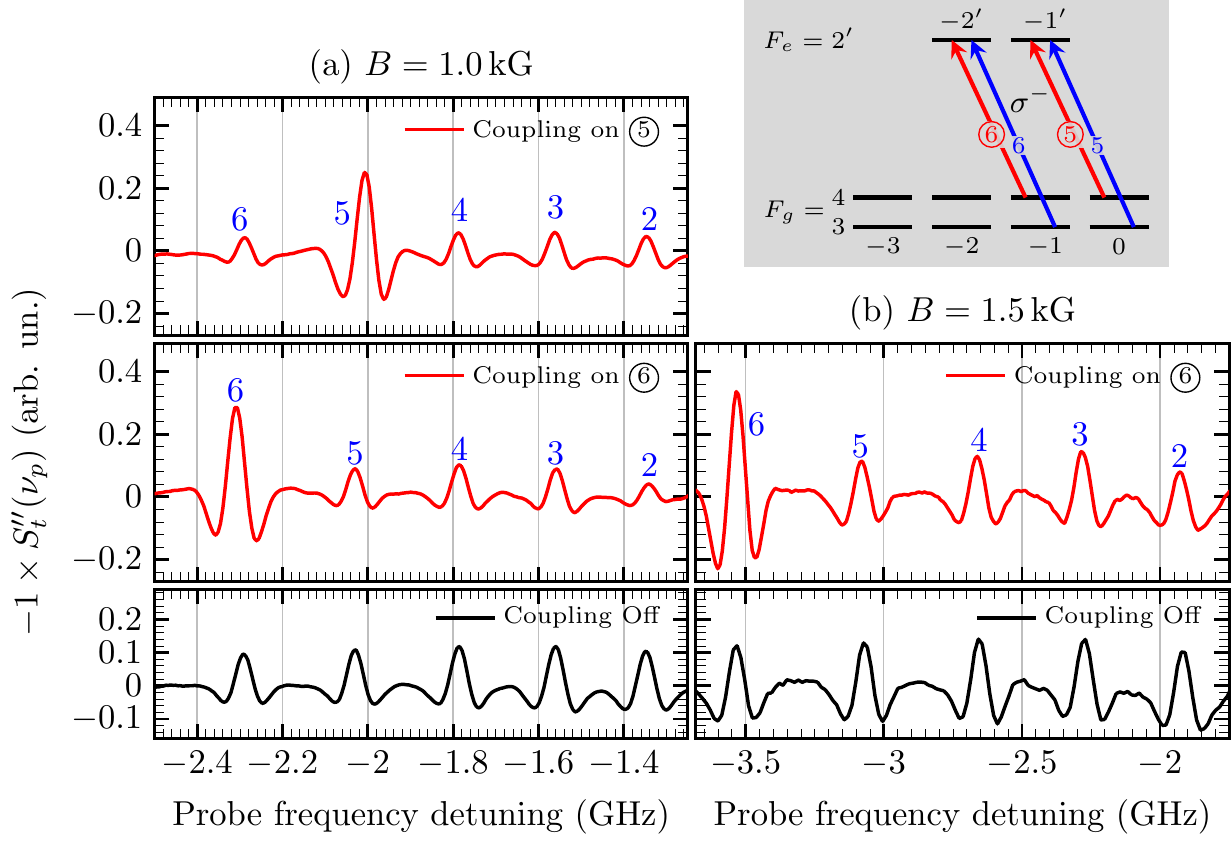}
    \caption{Probe second derivative spectra at two intermediate magnetic field values. (a)~$B=1.0\,\rm{kG}$, coupling tuned to \textcircled{5} (top), \textcircled{6} (middle), or switched-off (bottom); (b) $B=1.5\,\rm{kG}$, coupling tuned to \textcircled{6} (top) or switched-off (bottom). The zero frequency detuning is set to the unperturbed $3 \rightarrow  2'$ transition. The inset shows the $\Lambda$-systems involved in the realization of EIT.}
    \label{fig:SD}
\end{figure}

Figure\,\ref{fig:SD}(a) shows SD probe transmission spectrum in a longitudinal magnetic field of 1.0\,kG. The probe frequency is scanned in the vicinity of the transition 5 and 6, while the coupling frequency is resonant with MI transition \textcircled{5} (top panel) or transition \textcircled{6} (middle panel). The $\Lambda$-systems for the formation of EIT resonances using MI transitions \textcircled{5} and \textcircled{6} are shown in the right inset; both the probe and the coupling radiation must have $\sigma^-$ polarization for the EIT resonance to occur. The bottom panel shows SD probe transmission spectrum when the coupling radiation is switched-off.

Figure\,\ref{fig:SD}(b), shows SD of transmission spectrum at a field of $B= 1.5 \rm{kG}$ this time. The probe radiation frequency demonstrate higher transmission on transition 6 when the coupling frequency is resonant with the MI transition \textcircled{6}. As the probabilities of transitions $|3, -3\rangle \rightarrow |2', -2'\rangle$ and $|4, -3\rangle \rightarrow |2', -2'\rangle$  ($\sigma^+$ polarization) are close to zero they have been omitted on the diagram. The bottom panel shows the SD transmission spectrum in the absence of coupling radiation where only VSOP resonances can be seen.

From Figs.\,\ref{fig:TR} and \ref{fig:SD}, MI transitions \textcircled{4}, \textcircled{5} and \textcircled{6} allow the formation of EIT resonances when the coupling radiation is in resonance with these transitions. The generated EIT resonances can be strongly frequency-shifted by fine tuning the magnetic field in the range 1.0 to 2\,kG. It is indeed possible to carry out a shift down to $-6\,\rm{GHz}$ in the low-frequency region relative to the unperturbed $3\rightarrow2'$ transition, see Fig.\,\ref{fig:shift-proba}. It is important to emphasize again the following: as presented in Fig.\,\ref{fig:shift-proba}, the probabilities of transitions $3 \rightarrow 2'$, numbered 9 to 13 ($\sigma^+$ polarized radiation), rapidly tend to zero for $B > 100\,\rm{G}$. For this reason the formation of EIT resonances such as in Figs.\,\ref{fig:TR} and \ref{fig:SD} is only possible when the probe beam is $\sigma^-$-polarized, tuned in the vicinity of transitions 2 to 6, while the coupling beam has the same polarization. The comparison of EIT resonance $5$ in Figs.\,\ref{fig:TR}(a) and \ref{fig:SD}(a) shows that SD allows for a spectral narrowing by a factor about 1.6, demonstrating higher contrast. This is because the width of peaks in the SD spectrum strongly depends on the transitions spectral width in the original spectrum. Comparing to the formation of the EIT resonances at low fields reported on Cs D$_1$ line \cite{sargsyanJModOpt2015}, where seven EIT resonances equidistantly spaced in frequency were simultaneously formed, tuning directly on a single MI transition at intermediate magnetic fields allows to form only one EIT resonance at a time yet with much higher contrast. Indeed the amplitude of EIT resonances in \cite{sargsyanJModOpt2015} were rather small.

\section{Conclusion}

In conclusion, we have demonstrated the promise of using magnetically induced MI transitions $4 \rightarrow 2'$ of the Cs D$_2$ line at intermediate magnetic fields for the realization of EIT resonances. For the first time, and in contrast to \cite{sargsyanOptlett2019,sargsyanJETP2021}, MI transitions have been used as a coupling transition (with fixed frequency) to realize EIT. Three out of five of the $4 \rightarrow 2'$ MI transitions have been tuned on with the coupling laser to form a single EIT resonance in the spectrum of a probe beam scanned in the vicinity of the Cs D$_2$ line  $3\rightarrow2'$ transition perturbed by a magnetic field. This was achieved for fields ranging from 1.0 to 1.5\,kG and in a NC with a thickness $\ell= 852\, \rm{nm}$. We expect the generation of such EIT resonances to remain possible in the larger range 0.3 to 2.5\,kG as the probabilities of these three MI transitions are sufficiently high. 
In conjunction with what was observed in the weak-probe regime, the following rule of thumb was established: if MI transitions verifying the selection rule $\Delta F=\pm2$ are used for one of the two transitions of the $\Lambda$-system, then both probe and coupling radiation must have $\sigma^\pm$ circular polarization, respectively, to allow for the formation of EIT resonances. Thus, the generation of EIT resonances is affected by the circular dichroism induced by the presence of a magnetic field. 

Thanks to the small thicknesses accessible, NCs allow for the use of strong permanent magnets without largely broadening the resonances through field gradients. In large magnetic fields, the frequency shift of the EIT resonance can reach several GHz, which is of practical interest for the development of new frequency ranges in fields based on alkali vapor cells, for example for laser frequency stabilization or atomic Faraday filters operating at frequencies largely shifted from that of unperturbed atomic transitions, see e.g. \cite{mathewOL2018}. One can for example tune, in frequency and amplitude, the transparency of the vapor by either adjusting the magnetic field and/or tuning the coupling laser's frequency on the proper MI transition.
Note finally that the spectral width of the EIT resonances can be reduced by several orders of magnitude by using coherently coupled probe and coupling radiations and a centimeter cell filled with Cs vapor \cite{kitching2018chip}.

\section*{Acknowledgement}
The work was supported by the Science Committee of RA, in the frame of the research project No 21T-1C005. E.K. acknowledges support from the German Federal Ministry of Education and Research (BMBF) within the Quantentechnologien program (FKZ 13N15064).

\section*{References}
\bibliography{Bib-PhysLett2021}

\begin{thebibliography}{10}
\expandafter\ifx\csname url\endcsname\relax
  \def\url#1{\texttt{#1}}\fi
\expandafter\ifx\csname urlprefix\endcsname\relax\def\urlprefix{URL }\fi
\expandafter\ifx\csname href\endcsname\relax
  \def\href#1#2{#2} \def\path#1{#1}\fi

\bibitem{smith2004role}
D.~A. Smith, I.~G. Hughes, The role of hyperfine pumping in multilevel systems
  exhibiting saturated absorption, American Journal of Physics 72~(5) (2004)
  631--637.

\bibitem{scottoPRA2015}
S.~Scotto, D.~Ciampini, C.~Rizzo, E.~Arimondo, Four-level $\mathsf{N}$-scheme
  crossover resonances in {Rb} saturation spectroscopy in magnetic fields,
  Phys. Rev. A 92 (2015) 063810.
\newblock \href {http://dx.doi.org/10.1103/PhysRevA.92.063810}
  {\path{doi:10.1103/PhysRevA.92.063810}}.

\bibitem{tremblayPRA1990}
P.~Tremblay, A.~Michaud, M.~Levesque, S.~Th{\'e}riault, M.~Breton, J.~Beaubien,
  N.~Cyr, Absorption profiles of alkali-metal {D} lines in the presence of a
  static magnetic field, Phys. Rev. A 42~(5) (1990) 2766.

\bibitem{sargsyanApplPhysLet2008}
A.~Sargsyan, G.~Hakhumyan, A.~Papoyan, D.~Sarkisyan, A.~Atvars, M.~Auzinsh, A
  novel approach to quantitative spectroscopy of atoms in a magnetic field and
  applications based on an atomic vapor cell with l= $\lambda$, Applied Physics
  Letters 93~(2) (2008) 021119.

\bibitem{sargsyanLPL2014}
A.~Sargsyan, A.~Tonoyan, G.~Hakhumyan, A.~Papoyan, E.~Mariotti, D.~Sarkisyan,
  Giant modification of atomic transition probabilities induced by a magnetic
  field: forbidden transitions become predominant, Laser Phys. Lett. 11~(5)
  (2014) 055701.

\bibitem{scottoThesis2016}
S.~Scotto, Rubidium vapors in high magnetic fields, Ph.D. thesis,
  Universit{\'e} Paul Sabatier-Toulouse III (2016).

\bibitem{tonoyanEPL2018}
A.~Tonoyan, A.~Sargsyan, E.~Klinger, G.~Hakhumyan, C.~Leroy, M.~Auzinsh,
  A.~Papoyan, D.~Sarkisyan, Circular dichroism of magnetically induced
  transitions for {D}$_2$ lines of alkali atoms, Eur. Phys. Lett. 121~(5)
  (2018) 53001.

\bibitem{sargsyanJetpLett2021}
A.~Sargsyan, A.~Tonoyan, D.~Sarkisyan, Strongest magnetically induced
  transitions in alkali metal atoms, JETP Letters 113~(10) (2021) 605--610.

\bibitem{sargsyanPhysLetA2021}
A.~Sargsyan, A.~Amiryan, A.~Tonoyan, E.~Klinger, D.~Sarkisyan, Circular
  dichroism in atomic vapors: magnetically induced transitions responsible for
  two distinct behaviors, Physics Letters A 390 (2021) 127114.

\bibitem{auzinshBook2010}
M.~Auzinsh, D.~Budker, S.~Rochester, Optically polarized atoms: understanding
  light-atom interactions, Oxford University Press, 2010.

\bibitem{sarkisyanJETP2020}
D.~Sarkisyan, G.~Hakhumyan, A.~Sargsyan, “unmoved” atomic transitions of
  alkali metals in external magnetic fields, Journal of Experimental and
  Theoretical Physics 131~(5) (2020) 671--678.

\bibitem{zielinska2012ultranarrow}
J.~A. Zieli{\'n}ska, F.~A. Beduini, N.~Godbout, M.~W. Mitchell, Ultranarrow
  faraday rotation filter at the rb d 1 line, Optics letters 37~(4) (2012)
  524--526.

\bibitem{keaveney2018optimized}
J.~Keaveney, S.~A. Wrathmall, C.~S. Adams, I.~G. Hughes, Optimized ultra-narrow
  atomic bandpass filters via magneto-optic rotation in an unconstrained
  geometry, Optics letters 43~(17) (2018) 4272--4275.

\bibitem{fleischhauerRevModPhys77}
M.~Fleischhauer, A.~Imamoglu, J.~P. Marangos, Electromagnetically induced
  transparency: Optics in coherent media, Reviews of modern physics 77~(2)
  (2005) 633.

\bibitem{novikova2012electromagnetically}
I.~Novikova, R.~L. Walsworth, Y.~Xiao, Electromagnetically induced
  transparency-based slow and stored light in warm atoms, Laser \& Photonics
  Reviews 6~(3) (2012) 333--353.

\bibitem{yudin2010vector}
V.~Yudin, A.~Taichenachev, Y.~Dudin, V.~Velichansky, A.~Zibrov, S.~Zibrov,
  Vector magnetometry based on electromagnetically induced transparency in
  linearly polarized light, Physical Review A 82~(3) (2010) 033807.

\bibitem{belfi2007cesium}
J.~Belfi, G.~Bevilacqua, V.~Biancalana, S.~Cartaleva, Y.~Dancheva, L.~Moi,
  Cesium coherent population trapping magnetometer for cardiosignal detection
  in an unshielded environment, JOSA B 24~(9) (2007) 2357--2362.

\bibitem{PhysRevA.100.063427}
N.~Thaicharoen, K.~R. Moore, D.~A. Anderson, R.~C. Powel, E.~Peterson,
  G.~Raithel,
  \href{https://link.aps.org/doi/10.1103/PhysRevA.100.063427}{Electromagnetically
  induced transparency, absorption, and microwave-field sensing in a rb vapor
  cell with a three-color all-infrared laser system}, Phys. Rev. A 100 (2019)
  063427.
\newblock \href {http://dx.doi.org/10.1103/PhysRevA.100.063427}
  {\path{doi:10.1103/PhysRevA.100.063427}}.
\newline\urlprefix\url{https://link.aps.org/doi/10.1103/PhysRevA.100.063427}

\bibitem{holloway2021electromagnetically}
C.~L. Holloway, N.~Prajapati, J.~Kitching, J.~A. Sherman, C.~Teale,
  A.~Rufenacht, A.~B. Artusio-Glimpse, M.~T. Simons, A.~K. Robinson, E.~B.
  Norrgard, Electromagnetically induced transparency based rydberg-atom sensor
  for quantum voltage measurements, arXiv preprint arXiv:2110.02335.

\bibitem{vanier2005atomic}
J.~Vanier, Atomic clocks based on coherent population trapping: a review,
  Applied Physics B 81~(4) (2005) 421--442.

\bibitem{hafiz2020protocol}
M.~A. Hafiz, R.~Vicarini, N.~Passilly, C.~Calosso, V.~Maurice, J.~W. Pollock,
  A.~Taichenachev, V.~Yudin, J.~Kitching, R.~Boudot, Protocol for light-shift
  compensation in a continuous-wave microcell atomic clock, Physical Review
  Applied 14~(3) (2020) 034015.

\bibitem{sargsyanOptlett2019}
A.~Sargsyan, A.~Tonoyan, A.~Papoyan, D.~Sarkisyan, Dark resonance formation
  with magnetically induced transitions: extension of spectral range and giant
  circular dichroism, Optics letters 44~(6) (2019) 1391--1394.

\bibitem{sargsyanJETP2021}
A.~Sargsyan, A.~Tonoyan, D.~Sarkisyan, Application of magnetically induced
  transitions of the {$^{85}$Rb} {D$_2$} line in coherent processes, Journal of
  Experimental and Theoretical Physics 133~(1) (2021) 16--25.

\bibitem{vassilievRevSciInstrum2006}
V.~Vassiliev, S.~Zibrov, V.~Velichansky, Compact extended-cavity diode laser
  for atomic spectroscopy and metrology, Review of scientific instruments
  77~(1) (2006) 013102.

\bibitem{sargsyan2009efficient}
A.~Sargsyan, A.~Papoyan, D.~Sarkisyan, A.~Weis, Efficient technique for
  measuring laser frequency stability, The European Physical Journal Applied
  Physics 48~(2) (2009) 20701.

\bibitem{sargsyanJModOpt2015}
A.~Sargsyan, C.~Leroy, Y.~Pashayan-Leroy, S.~Cartaleva, D.~Sarkisyan,
  High-contrast dark resonances on the d 1 line in cesium nanocell: the
  advantages compared with the other alkali d lines, Journal of Modern Optics
  62~(10) (2015) 769--777.

\bibitem{andreevapPhysRevA2007}
C.~Andreeva, S.~Cartaleva, L.~Petrov, S.~Saltiel, D.~Sarkisyan,
  T.~Varzhapetyan, D.~Bloch, M.~Ducloy, Saturation effects in the sub-doppler
  spectroscopy of cesium vapor confined in an extremely thin cell, Physical
  Review A 76~(1) (2007) 013837.

\bibitem{sargsyan2012electromagnetically}
A.~Sargsyan, C.~Leroy, Y.~Pashayan-Leroy, D.~Sarkisyan, D.~Slavov,
  S.~Cartaleva, Electromagnetically induced transparency and optical pumping
  processes formed in cs sub-micron thin cell, Optics Communications 285~(8)
  (2012) 2090--2095.

\bibitem{sargsyanOL2019}
A.~Sargsyan, A.~Amiryan, Y.~Pashayan-Leroy, C.~Leroy, A.~Papoyan, D.~Sarkisyan,
  Approach to quantitative spectroscopy of atomic vapor in optical nanocells,
  Opt. Lett. 44~(22) (2019) 5533--5536.
\newblock \href {http://dx.doi.org/10.1364/OL.44.005533}
  {\path{doi:10.1364/OL.44.005533}}.

\bibitem{kitching2018chip}
J.~Kitching, Chip-scale atomic devices, Applied Physics Reviews 5~(3) (2018)
  031302.

\bibitem{bhushanPhysRevA2019}
S.~Bhushan, V.~S. Chauhan, M.~Dixith, R.~K. Easwaran, Effect of magnetic field
  on a multi window ladder type electromagnetically induced transparency with
  87rb atoms in vapour cell, Physics Letters A 383~(31) (2019) 125885.

\bibitem{savitzkyAnChem1964}
A.~Savitzky, M.~J. Golay, Smoothing and differentiation of data by simplified
  least squares procedures., Analytical chemistry 36~(8) (1964) 1627--1639.

\bibitem{talsky1994}
G.~Talsky, Derivative spectrophotometry: low and high order, VCH Publishers,
  1994.

\bibitem{mathewOL2018}
R.~S. Mathew, F.~Ponciano-Ojeda, J.~Keaveney, D.~J. Whiting, I.~G. Hughes,
  Simultaneous two-photon resonant optical laser locking ({STROLL}ing) in the
  hyperfine paschen--back regime, Opt. Lett. 43~(17) (2018) 4204--4207.
\newblock \href {http://dx.doi.org/10.1364/OL.43.004204}
  {\path{doi:10.1364/OL.43.004204}}.

\end{thebibliography}

\end{document}